# Ion energy balance during fast wave heating in TORE SUPRA

T. Hutter, A. Bécoulet, J.P. Coulon, B. Saoutic, V. Basiuk, G. T. Hoang
Association Euratom-CEA sur la Fusion
Centre d'Etudes de Cadarache
13108 Saint Paul Lez Durance (FRANCE)

**Introduction**

Direct coupling of the fast magnetosonic wave to the electrons has been recently studied on TORE SUPRA. Preliminary experiments were dedicated to optimise the scenario for Fast Wave Electron Heating (FWEH) and Current Drive (FWCD). In a first part, thermal kinetic and diamagnetic energy are compared when fast wave is applied to the plasma in two different regimes: 1/ the minority hydrogen heating scenario (ICRH), 2/ the direct electron damping. Effects of ion resonant layers, marginally present in the plasma in the later regime (FWEH), is then presented and discussed.

In the following, all plasmas are limited on the carbon inner wall, with $R = 2.28$ m and $a = 0.72$ m, and, unless specified, are helium plasmas with a small hydrogen concentration ($n_H/n_e < 5\%$).

**Thermal energy content**

On TORE SUPRA, time resolved energy spectra of fast hydrogen and deuterium neutrals are routinely measured by charge exchange (CX) analysers [1] of the E∥B type (46 channels per mass, H and D, Emax = 300 keV) with different lines of sight: five in a poloidal cross-section having a tangency radius $R_t = 0.43$ m (near "perpendicular"), and one in the equatorial plane with $R_t = 1.9$ m (near "parallel" to the magnetic axis, see figure 4).

Concerning the ion energy balance, the ion temperature profiles are simulated including: 1/ a best fit to the passive measured spectra of each analyser, 2/ for all the analysers, the same normalisation factor for the neutral density profile (fast computed consistently with ion temperature[2]) and 3/ the consistency with neutron yield and plasma composition. Standard diagnostics provide the other experimental data required in the simulation code ($n_e(r)$, $T_e(r)$, $Z_{eff}$, neutron yield and plasma geometry).

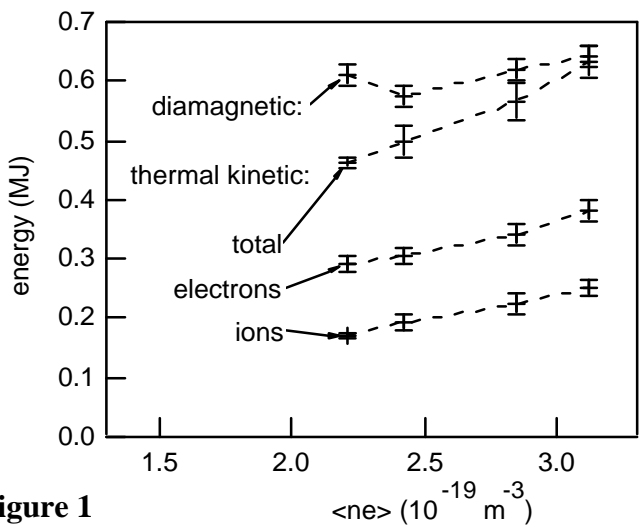

**Figure 1**

A 4-step electron density scan is performed on a 4 s, 4 MW ICRH pulse ($f = 57.4$ MHz, dipole phasing), launched in a mainly deuterium plasma ($n_D/n_{He}$ ≈ 2, B = 3.9 T, Ip=1.3 MA). Electron and ion kinetic thermal energies are shown on figure 1. In the ohmic phase, diamagnetic and kinetic energy are found in good agreement. The lowest density point corresponds to a monster sawtooth. For the other points, data are averaged during 100 ms (about two normal sawteeth). If we assume that perpendicular supra-thermal ion energy content is the difference between diamagnetic and thermal kinetic energy, it is within the error bars for the high density point, but clearly increases with decreasing density. The thermal energy content has also been evaluated on a 4 MW pulse of FWCD ($f = 47.7$ MHz, co-current phasing) launched in mainly helium plasma ($n_D/n_{He}$ ≈ 0.7, B = 2.2 T, Ip = 0.76 MA). It is found equal to the diamagnetic energy, indicating that the supra-thermal particle energy content, if present, is negligible.

## Hydrogen spectra during FWEH optimisation

To get a better efficiency for FWEH/CD, it is necessary to avoid any ion damping of the wave. Due to the outgasing of the TORE SUPRA carbon inner wall, it is impossible to get rid of hydrogen and deuterium. A precise monitoring of the position of the ion cyclotron layers is thus a necessity. The CX analysers, both parallel and perpendicular, appear to be the best tool to optimise the scenarios. For f=47.7 MHz (dipole phasing, 2MW, 2s), the magnetic field B has been scanned between 2.10 and 2.27 T on four consecutive shots (Ip = 0.75MA). The first harmonic hydrogen layer, $\Omega_{CH}$, is on the high field side (HFS) of the discharge, the second one, $2\Omega_{CH}$, on the low field side (LFS), and the third harmonic of majority ions (deuterium or helium, $3\Omega_{CD}$) is at the plasma centre. In the present plasma conditions, no significant damping on $3\Omega_{CD}$ is expected. This is experimentally confirmed by the usual maxwellian shape of the observed deuterium spectra (figure 6). Hydrogen neutral spectra measured during these shots (analysers time resolution: 65 ms) are shown on figure 2. One can notice that the perpendicular tail vanishes when increasing B from 2.1 to 2.24 T, and that above 2.2 T a large non maxwellian tail appears on the parallel analyser.

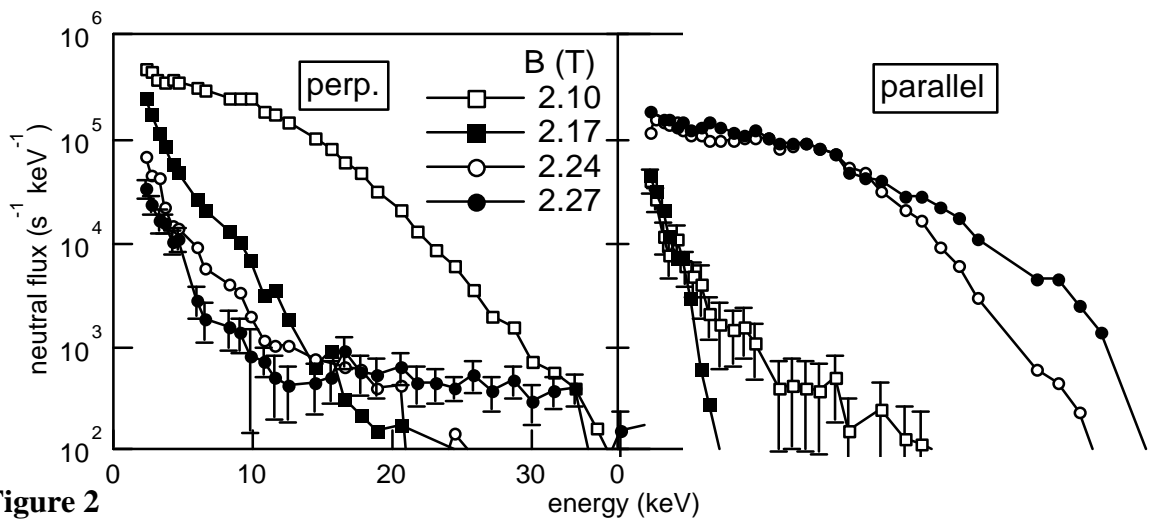

**Figure 2**

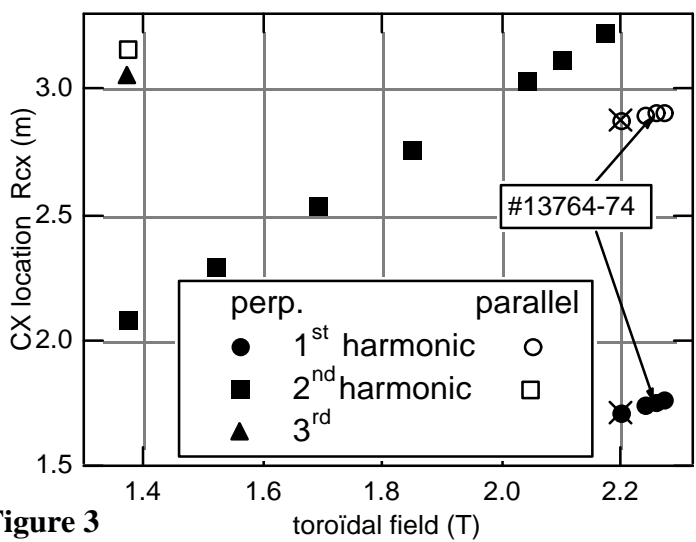

**Figure 3**

An ion of velocity v directed towards the CX analyser (and of parallel velocity $v_{//}$), making a CX at major radius Rcx, satisfies $Rt/Rcx = v_{//} / v$. According to the adiabatic theory of particle trajectories in tokamak, this ion, before the CX, can cross a cyclotron resonant layer ($n\Omega_{CH}$, n=1,2,...) only if $Rcx - Rt^2 / Rcx \leq R_{n\Omega_{CH}}$. The equality corresponds to a trapped ion with banana tips on the resonant layer, which interacts strongly with the wave. The corresponding value of Rcx is shown on figure 3 for each harmonic layer present in the plasma ($1.56 < R_{n\Omega_{CH}} < 3.2$). One can check that, in the present B-scan, particles interacting with the $\Omega_{CH}$ layer can be detected by the parallel analyser when doing their CX on the LFS and by the perpendicular analyser when doing their CX on the HFS. However the energy of the ions in the present case is low

(E<20keV, cf. figure 2). The particles making their CX on the HFS are thus mostly re-ionised before reaching the perpendicular analyser. The ions accelerated on $\Omega_{CH}$ (HFS) are then mainly detected by the parallel analyser. And in the same way, the perpendicular analyser mainly detects the ions accelerated on the $2\Omega_{CH}$ (LFS) layer. It is thus possible, using both CX analysers, to explain the B-scan experiments results as the $2\Omega_{CH}$ layer moving outside the plasma, and the $\Omega_{CH}$ layer progressively entering on the HFS. Finally B=2.17 T represents the optimum field, where no significant ion damping competes with the direct electron heating.

This is also well confirmed by the expected positions of the cyclotron layers shown in Table I, except that at B=2.10T the $2\Omega_{CH}$ layer is 4 cm outside the plasma and that a significant tail is however observed. One must remember here that the ripple effect is quite high on the low field side of TORE SUPRA. Indeed, for the case B = 2.10T, R+a increases by 3 cm and $R_{2\Omega_{CH}}$ decreases by 15 cm between two coils[3]. So it is clear on figure 4 that confined protons can interact with the wave.

**table I**: relative position (in meter) of the harmonic resonant layers. A negative value indicate a layer outside the plasma.

| B(T) | $R_{\Omega_{CH}}$-(R-a) | R+a-$R_{2\Omega_{CH}}$ |
|------|-------------------------|------------------------|
| 2.10 | -.04                    | -.04                   |
| 2.17 | .01                     | -.14                   |
| 2.24 | .06                     | -.24                   |
| 2.27 | .08                     | *(-.29)                |

*: no signification (out of the torus).

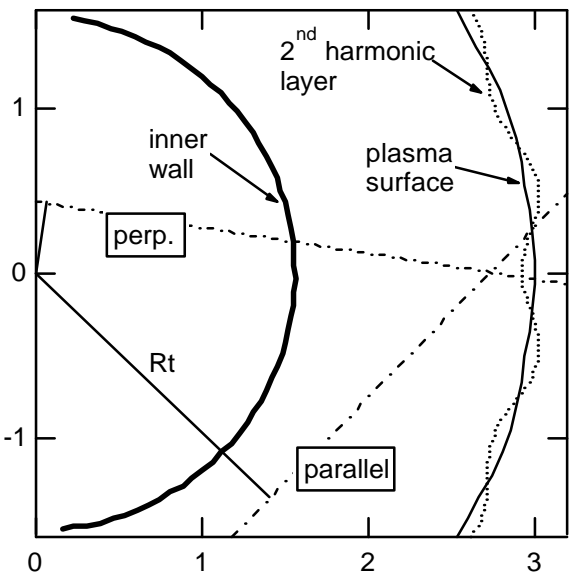

**Figure 4**: *lines of sight of the CX analysers in the equatorial plane.*

The Doppler effect also broadens the cyclotron layer, thus being coupled to the ions a few centimetres deeper in the plasma. Finally, since few obstacles are located on the outer part of the plasma chamber, one can expect sufficient confinement of 5-30 keV protons in the low electron, high neutral density plasma located on the low field side of the torus, thus giving rise to that relatively high neutral signal observed on the perpendicular analysers when 2.1 < B < 2.2 T ($2\Omega_{CH}$, upper right black square on figure 3).

## Ion-electron competitions during FWEH/CD experiments

A larger B-scan has been performed (1.37<B(T)<2.2), at f=47.7MHz, in order to investigate the possibility for maximizing the electron single-pass absorption (which scales as $B^{-3}$). But lowering B also means entering the $2\Omega_{CH}$ layer into the plasma, and thus a competition with ion damping (which increase also with ion temperature Ti). Figure 5 displays the CX analysers measurements for various B values, the corresponding cyclotron layers locations being given in Table II. One clearly distinguish the increasing absorption on the $2\Omega_{CH}$ layer as it enters the plasma towards the centre (and experiences an increasing Ti, B from 2.2 to 1.69T), followed by a decrease as B varies from 1.69 to 1.37T. The FW single-pass absorption computations, as well as full-wave simulations, confirm the observations, revealing a switch between the electron and ion dampings when the $2\Omega_{CH}$ layer is at the plasma centre (B=1.69T).

Another example of such a competition between ions and electrons has been observed when phasing the currents between the FW antenna straps, for FWCD experiments. In that case, the averaged parallel wave number decreases, lowering the electron single-pass absorption. This is well confirmed by the neutral spectra measurements, in cases where the set f / B is not optimised: for 48MHz / 2.25T, $\Omega_{CH}$ layer is slightly inside the plasma on the high field side. On figure 6, hydrogen spectra measured during FWEH for two different shots, with these f / B values, are presented. We can see that spectra remain low in dipole phasing (#13764), but with the co-current phasing (#13774), we observe a large increase of the parallel analyser signal and the appearance of a "flat" spectra on the perpendicular one for energy above 20 keV (fast neutral coming from HFS and crossing the plasma, despite higher density than in #13764!).

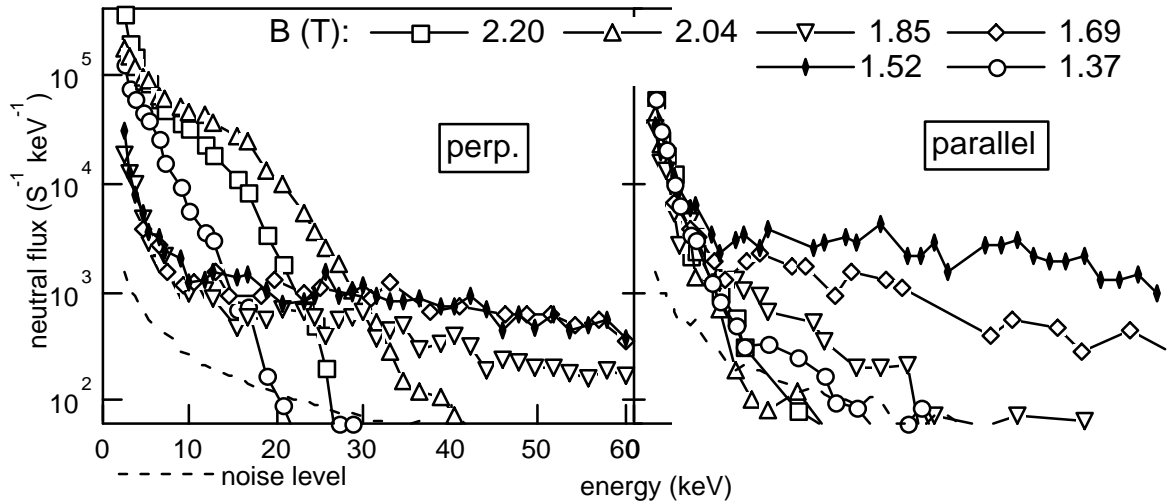

**Figure 5**

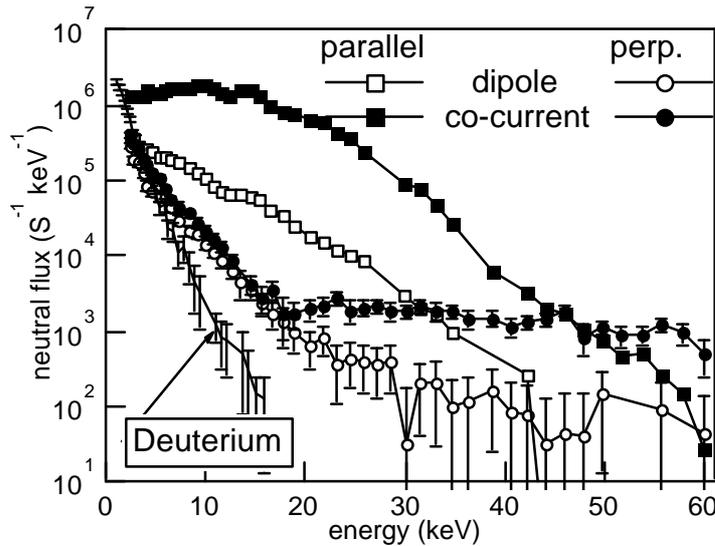

**Figure 6**

**table II**: $R_m$ : magnetic axis radius
($I_p$ = 0.4 MA, $P_{RF}$ = 1.5 MW)

| B(T) | $R_{\Omega_{CH}}$-(R-a) | $R_{2\Omega_{CH}}$-$R_m$ | (R+a)-$R_{3\Omega_{CH}}$ |
|---|---|---|---|
| 2.20 | .05 | .81 | * |
| 2.04 | *(-.17) | .59 | * |
| 1.85 | * | .31 | * |
| 1.69 | * | .06 | * |
| 1.52 | * | -.18 | *(-.31) |
| 1.37 | * | -.39 | 0 |

*: no signification (out of the torus).

## Conclusions

On TORE SUPRA CX analysers 1/ prove to be a powerful tool in the determination and optimisation of the FWEH/CD scenarios, 2/ permit direct observation of competing damping mechanisms for ions and electrons.

---

[1] similar to those developed for TFTR: A. L. Roquemore et al., Rev. Sci. Instrum. 56, 1120 (1985).
[2] H. Capes, P. Laporte, T. Hutter and J-C. M. de Haas, this conference.
[3] R. Arslanbekov, Y. Peysson et al, this conference.